\documentclass[12pt,showpacs,aps]{revtex4}
\def\comment#1{}

\begin{document}

\preprint{}

\title{Spontaneous photon emission from a semiconductor}

\author{She-Sheng Xue}

\email{xue@icra.it}

\affiliation{ICRA and
Physics Department, University of Rome ``La Sapienza", 00185 Rome, Italy}



\begin{abstract}

Semiconductor's energy band and its degeneracy are altered when an external
magnetic field varies, analogously to the appearance of Landau levels.
Such alternation leads to the variation of energy levels and thus quantum-mechanical instability. 
As a consequence, spontaneous photon emission and/or paramagnetically screening effect
take place. By computing total energy variation, we discuss why these effects are
bound to occur. In addition, we calculate the rate and spectrum of spontaneous 
photon emission, which can be experimentally tested.
 
\end{abstract}


\maketitle

\section{Introduction}\label{int}

semiconductors are characterized by energy gaps $\Delta$ between the top 
of the highest filled valence band(s), where all electron states are occupied, and 
the bottom of the lowest empty conduction band(s), where all electron states are 
unoccupied at zero temperature $T=0$. When the temperature is not zero and below the melting point, there is a
non vanishing probability that some electrons will be thermally excited across
energy gaps $\Delta$ into the lowest unoccupied conducting bands, and therefore some holes 
will be left behind in the valence band. The probability is roughly proportional 
to $e^{-\Delta/k_BT}$, where $k_B$ is the Boltzmann constant. At room 
temperature $k_BT\simeq 0.024$eV, energy gaps $\Delta\sim 10^{-1}$eV that 
thermal excitation can lead to observable conductivity. The vacuum in quantum field theories has similar 
characteristics: (i) the negative energy spectrum, 
as valence band(s), in which all electron states are occupied; (ii) 
the positive energy spectrum, as conducting band(s), in which all electron states 
are unoccupied; (iii) the energy gap $2m_ec^2\simeq 1.04$MeV separating
the positive energy spectrum from the negative energy spectrum. However, thermal 
excitations are almost impossible for $k_BT\ll 1.04$MeV. Thus, a semiconductor at 
temperature $k_BT\ll \Delta$ described by quantum mechanics is very much analogous 
to the vacuum described by quantum field theories.      

In Refs. \cite{xue}, we studied and presented effects that when 
an external magnetic field is turned on, the vacuum of quantum field theories is unstable and bound to 
decay, leading to magnetic screen and/or spontaneous photon emission.
It is difficult to detect such effects, since the vacuum has a huge volume and 
the relaxation time is small. 
Due to the similarity between vacuum and semiconductor mentioned in above, we extend the analysis in 
the vacuum case  \cite{xue} to semiconductor case. In the framework of quantum mechanics, 
we calculate the energy variation of semiconductor due to the presence of magnetic field,
and discuss why such energy variation leads to quantum mechanical instability and spontaneous photon emission.
We calculate the rate and spectrum of spontaneous 
photon emission, which can be experimentally tested.
  
\section{Bloch electron spectrum and state}

The problem of electrons in a semiconductor is in principle a many-electron problem, 
for the full Hamiltonian of the semiconductor contains not only the one-electron potentials 
describing the interactions of the electrons with massive atomic nuclei 
arranged in a periodic array, but also
pair potentials describing the electron-electron interactions. In the independent electron 
approximation these interactions are represented by an effective potential $U({\bf r})$ 
with the periodicity of the underlying Bravais lattice of the spacing vector ${\bf a}$, the effective Hamiltonian
is given by
\begin{equation}  
H_{\rm eff}= -\frac{\hbar^2}{2m}({\bf \nabla })^2+U({\bf r});\quad U({\bf r}+{\bf a})=U({\bf r}).
\label{period}
\end{equation}
The lattice spacing $|{\bf a}|$, which is the scale of periodicity of the potential, 
is about $10^{-8}$cm. At this typical scale, it is essential to use quantum mechanics 
in accounting for the effect of the periodic potential (\ref{period}) on electronic motion. 
Independent electrons, each of which obeys a one-electron Schr\"odinger equation with the Hamiltonian (\ref{period}), 
are known as {\it Bloch electrons},
\begin{equation}  
H_{\rm eff}\psi_{{\mathcal N}{\bf k}}({\bf r})={\mathcal E}_{{\mathcal N}{\bf k}}\psi_{{\mathcal N}{\bf k}}({\bf r});\quad \psi_{{\mathcal N}{\bf k}}({\bf r},t)=\exp\left(-\frac{i}{\hbar}{\mathcal E}_{{\mathcal N}{\bf k}}t\right)
\psi_{{\mathcal N}{\bf k}}({\bf r}),
\label{periodstate0}
\end{equation}
where ${\bf k}$ and ${\mathcal E}_{{\mathcal N}{\bf k}}$ are Bloch electron's momentum and 
energy-spectrum in energy band ${\mathcal N}$. 
The corresponding Bloch electron states $\psi_{{\mathcal N}{\bf k}}({\bf r})$ obey Bloch's theorem,
\begin{equation}  
\psi_{{\mathcal N}{\bf k}}({\bf r})=e^{\frac{i}{\hbar}{\bf k}\cdot{\bf r}}u_{{\mathcal N}{\bf k}}({\bf r}),
\quad u_{{\mathcal N}{\bf k}}({\bf r}+{\bf a})=u_{{\mathcal N}{\bf k}}({\bf r}),
\label{periodstate}
\end{equation}
where the periodic wave-function $u_{{\mathcal N}{\bf k}}({\bf r})$ is a solution to the differential equation,  
\begin{equation}  
\left[\frac{1}{2m}(\frac{\hbar}{i}{\bf \nabla }+ {\bf k})^2+U({\bf r})\right]
u_{{\mathcal N}{\bf k}}({\bf r})={\mathcal E}_{{\mathcal N}{\bf k}}u_{{\mathcal N}{\bf k}}({\bf r}).
\label{periodstateu}
\end{equation}
In principle, one can try to solve this eigenvalue problem and find      
Bloch electron states $\psi_{{\mathcal N}{\bf k}}({\bf r})$ and energy-spectrum 
${\mathcal E}_{{\mathcal N}{\bf k}}$ for different energy bands ``${\mathcal N}$''. Throughout 
the following discussions we consider a single energy band and drop explicit
reference to the band index ${\mathcal N}$. 

In a given energy band ${\mathcal N}$, we are
only interested in Bloch electron states $\psi_{\bf k}({\bf r})$ and energy-spectrum
${\mathcal E}_{\bf k}$ for small momenta $\bf k$ in the first Brillouin zone, 
$a_i k_i\ll \hbar, (i=x,y,z)$. In the momentum phase space (${\bf k}$-space), 
the Bloch electron energy-spectrum ${\mathcal E}_{\bf k}$ in the neighborhood of 
its maxima can be approximately expanded by the quadratic forms of small momenta $k_i$
\cite{solidbook}
\begin{equation}  
{\mathcal E}_{\bf k}\simeq {\mathcal E}_v- 
\left(\frac{k^2_x}{2m_x}+\frac{k^2_y}{2m_y}+\frac{k^2_z}{2m_z}\right),
\label{spectrum1}
\end{equation}
where ${\mathcal E}_v$ is the energy at the top of the valence band and 
$m_i, (i=x,y,z)$ determining the dynamics of Bloch's electron states near band maxima
are ``effective masses''. In Eq. (\ref{spectrum1}),
we have taken the origin of ${\bf k}$-space to lie at the band maximum and a
set of orthogonal principle axes $(\hat{\bf x},\hat{\bf y},\hat{\bf z})$. 
The constant-energy surface ${\mathcal E}_{\bf k}={\rm const.}$ is in general 
an ellipsoid. In the case of an axial symmetry w.r.t. the $\hat{\bf k}_z$ direction in the ${\bf k}$-space,
$m_x=m_y=m^*$ and  Eq. (\ref{spectrum1}) becomes  
\begin{equation}  
{\mathcal E}_{\bf k}\simeq {\mathcal E}_v- 
\left(\frac{k^2_x+k^2_y}{2m^*}+\frac{k^2_z}{2m_z}\right)={\mathcal E}_v- 
\left(\frac{A({\mathcal E}_{\bf k},k_z)}{2\pi m^*}+\frac{\bar k^2_z}{2m^*}\right),
\label{spectrum2}
\end{equation}
where $\bar k_z=(m^*/m_z)^{1/2}k_z$, $A({\mathcal E}_{\bf k},k_z)\equiv\pi(k^2_x+k^2_y)$ 
an cross-sectional area of energy surface ${\mathcal E}_{\bf k}={\rm const.}$ in a plane $k_z={\rm const.}$ 
normal to the $\hat{\bf k}_z$ direction in the $\bf k$-space. $m^*$ is a cyclotron effective mass,
\begin{equation}  
m^*({\mathcal E}_{\bf k},k_z)=-\frac{1}{2\pi}
\frac{\partial A({\mathcal E}_{\bf k},k_z)}{\partial{\mathcal E}_{\bf k}}.
\label{m*}
\end{equation}
Correspondingly, the Bloch electron state can be expressed as,
\begin{equation}  
\psi_{\bf k}({\bf r})=e^{\frac{i}{\hbar}(k_xx+k_yy+\bar k_zz)}u_{\bf k}({\bf r});
\quad \psi_{\bf k}({\bf r},t)=\psi_{\bf k}({\bf r})
e^{-\frac{i}{\hbar}{\mathcal E}_{\bf k}t}.
\label{periodstate1}
\end{equation} 

A wealth of important information on the electronic properties of metals and semiconductors comes 
from the measurements of their response to various probes in the 
presence of a uniform magnetic field $H$. In the case of free electrons in a uniform
magnetic field $H$ pointing to the $\hat {\bf z}$-direction, the Landau energy-levels are given by \cite{landau}
\begin{equation}  
{\mathcal E}^{\rm free}_n(k_z)=\frac{k^2_z}{2m} 
+(n+\frac{1}{2}+\sigma)\hbar\omega,\quad \omega=\frac{eH}{mc},
\label{spectruml}
\end{equation}
where $m$ is the electron mass, the quantum number $n=0,1,2,3,\cdot\cdot\cdot$ is associated with 
harmonic oscillations in the plane orthogonal to
${\bf H}$ and $\sigma=\pm 1/2$ for spin-$1/2$ particle. 
The Onsager generation of the Landau levels (\ref{spectruml}) to the energy band (\ref{spectrum1}) of 
Bloch electron case can be written as, 
\begin{equation}  
{\mathcal E}_n(k_z)={\mathcal E}_v- \left[\frac{k^2_z }{2m_z}
+(n+\lambda_o)\hbar\omega^*\right],\quad \omega^*=\frac{eH}{m^*c},
\label{spectrumh}
\end{equation}
where $\lambda_o$ is a constant independent of integer $n$. 
The cross-sectional area $A({\mathcal E}_{\bf k},k_z)$ (\ref{spectrum2}) 
is quantized,
\begin{equation}
A({\mathcal E}_{\bf k},k_z)=(n+\lambda)\Delta A,\quad \Delta A= \frac{2\pi\hbar eH}{c},
\label{quantum}
\end{equation}
where $\Delta A$ is the cross-sectional area 
corresponding to each quantum number $n$.

Since the magnetic field $\bf H$ is in the $\hat {\bf z}$ direction, we chose electromagnetic potential
$A_x=-Hy$ and $A_y=A_z=0$. The operator $(\frac{i}{\hbar}\nabla)^2$ in Eqs. (\ref{period},\ref{periodstateu}) 
will be replaced by 
$(\frac{i}{\hbar}\nabla-\frac{e}{c}A_x)^2$. This implies that 
the periodicity of the Hamiltonian (\ref{period}) and Bloch electron states 
(\ref{periodstate0},\ref{periodstateu}) is lost in the $\hat {\bf y}$ direction. 
Analogously to Eq. (\ref{periodstate1}), Bloch electron states in such 
external magnetic field are expressed in the form
\begin{equation}  
\psi^h_{\bf k}({\bf r})=e^{\frac{i}{\hbar}(k_xx+\bar k_zz)}\chi(y)u^h_{\bf k}({\bf r});
\quad \psi^h_{\bf k}({\bf r},t)=\psi^h_{\bf k}({\bf r})
e^{-\frac{i}{\hbar}{\mathcal E}_n(k_z)t}
\label{periodstateh}
\end{equation}
where the wave-function $u^h_{\bf k}({\bf r})$ is periodic only in $\hat {\bf x}$ and 
$\hat {\bf z}$ 
directions and unknown function $\chi(y)$ is no longer $\exp(\frac{i}{\hbar}k_yy)$ 
as in Eq. (\ref{periodstate1}). 

In the absence and presence of magnetic field, we only consider 
Bloch electron states $\psi_{\bf k}({\bf r})$ (\ref{periodstate1}) 
and $\psi^h_{\bf k}({\bf r})$ (\ref{periodstateh}), whose momenta $\bf k$ are in the first Brillouin zone, 
$|{\bf k}||{\bf a}|\ll \hbar$, corresponding to their energies 
${\mathcal E}_{\bf k}$ (\ref{spectrum1}) and ${\mathcal E}_n(k_z)$ 
(\ref{spectrumh}). 
The wavelengths of such Bloch electron states $\lambda=\hbar/|{\bf k}|$ are much 
larger than the lattice spacing $|{\bf a}|$. Within the range of wave-lengths $\lambda$, 
the wave-functions $u_{\bf k}({\bf r})$ (\ref{periodstate1}) and $u^h_{\bf k}({\bf r})$ (\ref{periodstateh}), 
whose the periodicity $|{\bf a}|\ll\lambda$, vary slowly and can be approximately treated as constants, 
\begin{equation}  
u_{\bf k}({\bf r})\simeq C;\quad u^h_{\bf k}({\bf r})\simeq C^h,
\label{periodstateapp}
\end{equation}
in such ``continuum limit'' $|{\bf k}||{\bf a}|\rightarrow 0$.

Given the volume of a semiconductor sample $V=L_x L_y L_z$, dividing the 
phase space $L_xL_y\Delta A$ by the volume $(2\pi \hbar)^2$ for a quantum state, 
one obtains the degeneracy of Hamiltonian eigenstates corresponding to 
each quantum number $n$ in Eqs. (\ref{spectruml},\ref{spectrumh}),
\begin{equation}
{\mathcal D}=2\frac{L_xL_y\Delta A}{(2\pi \hbar)^2}=2 \frac{L_xL_y eH}{2\pi\hbar c},
\label{degeneracy}
\end{equation}
including the factor of 2 for spin degeneracy. Note that the degeneracy ${\mathcal D}$ is the same in 
the Landau case for free electrons and the Onsager case for Bloch electrons. This degeneracy
reflects the fact that electron with a given energy ${\mathcal E}_{\bf k}$ and $k_z$ 
spirals about a line parallel to the magnetic field direction, which can have arbitrary
$x$- and $y$-coordinates. Since 
\begin{equation}
\frac{2 e}{2\pi\hbar c}\simeq 4.84\cdot 10^6 \frac{1}{{\rm cm}^2{\rm Gauss}},
\label{degeneracy1}
\end{equation}
in the field of a kilogauss and a sample 1 cm on a side, this degeneracy 
${\mathcal D}$ is about $10^{10}$.

\section{Energy gain and instability}

Integrating Eq. (\ref{spectrum2}) over all states in the phase space, 
one obtains the total energy of the semiconductor sample in absence of magnetic field $H$,
\begin{equation}  
E_{\rm tot}=2V\int\frac{dk_x dk_y}{(2\pi\hbar)^2}\int\frac{dk_z}{(2\pi\hbar)}
\left[{\mathcal E}_v- 
\left(\frac{k^2_x+k^2_y}{2\pi m^*}+\frac{k^2_z}{2m_z}\right)\right].
\label{energy}
\end{equation}
Analogously, the total energy of the semiconductor sample in presence of 
magnetic field $H$ yields,
\begin{equation}  
E^h_{\rm tot}=2V\frac{\Delta A}{(2\pi\hbar)^2}\sum_n\int\frac{dk_z}{2\pi\hbar}\left\{{\mathcal E}_v- \left[(n+\lambda_0)\hbar\omega^*+\frac{k^2_z}{2m_z}\right]\right\}.
\label{energyh}
\end{equation}
The factor 2 in Eqs. (\ref{energy},\ref{energyh}) accounts for spin degeneracy and the degeneracy (\ref{degeneracy}) has been taken into account in Eq. (\ref{energyh}).
There is a difference between the total energy (\ref{energy}) and the total energy 
(\ref{energyh}), i.e. an energy variation of the semiconductor sample when the magnetic 
field $H$ is turned on.  

In order to compute this energy difference, we rewrite the phase-space integration 
in (\ref{energy}) 
by using Eq. (\ref{quantum}) for continuous values of $n$,
\begin{equation}  
\int\frac{dk_x dk_y}{(2\pi\hbar)^2}=2\pi\int \frac{k_\perp dk_\perp}{(2\pi\hbar)^2}
=\frac{\Delta A}{(2\pi\hbar)^2} \int dn,
\label{intkn}
\end{equation}
where
\begin{equation}  
k_\perp\equiv\sqrt{k^2_x+k^2_y}=\sqrt{\frac{\Delta A}{\pi}}(n+\lambda_0)^{1/2}; \quad
dk_\perp=\frac{1}{2}\sqrt{\frac{\Delta A}{\pi}}(n+\lambda_0)^{-1/2}dn.
\label{kn}
\end{equation}
Thus Eq. (\ref{energy}) becomes
\begin{equation}  
E_{\rm tot}=2V\frac{\Delta A}{(2\pi\hbar)^2}\int dn \int\frac{dk_z}{2\pi\hbar}\left[{\mathcal E}_v- 
\left(\frac{(n+\lambda_0)\Delta A}{2\pi m^*}+\frac{k^2_z}{2m_z}\right)\right]=\int dn F(n),
\label{energyn}
\end{equation}
where we define the function $F(n)$
\begin{equation}  
F(n)\equiv 2V\frac{\Delta A}{(2\pi\hbar)^2}\int\frac{dk_z}{2\pi\hbar}
\left[{\mathcal E}_v- 
\left(\frac{(n+\lambda_0)\Delta A}{2\pi m^*}+\frac{k^2_z}{2m_z}\right)\right].
\label{fn}
\end{equation}
The range of integration over $k_z$ is approximately equal to $[-m_zc,+m_zc]$, 
where is the validity for the quadratic expansion (\ref{spectrum1}).   
Using the Euler-MacLaurin formula,
\begin{equation}
F(0)+F(1)+F(2)+\cdot\cdot\cdot - \int dn F(n)
=-\frac{1}{2!}B_2F'(0)-\frac{1}{4!}B_4F^{'''}(0)-\cdot\cdot\cdot,
\label{emformula}
\end{equation}
and Bernoulli numbers $B_2=1/6,B_4=-1/30,\cdot\cdot\cdot$, we obtain
the energy difference: 
\begin{equation}
E^h_{\rm tot}-E_{\rm tot}=
-\frac{1}{2!}B_2F'(0)\simeq - V\frac{\alpha}{3\pi}H^2 \left |\frac{m_z}{m^*}\right|,
\label{delta2}
\end{equation}
where the fine structure constant $\alpha\equiv e^2/(4\pi\hbar c)$. 
We need to sum over all contributions from different energy bands at different maxima in Eq. (\ref{delta2}). 

The energetic difference Eq. (\ref{delta2}) between the semiconductor states ($H=0$)
and ($H\not=0$) is negative, indicating that the total energy Eq. (\ref{energyh}) 
($H\not=0$) is smaller than the total energy Eq. (\ref{energy}) ($H=0$), i.e., 
the semiconductor state gains energy when the external magnetic field 
is turned on. The reasons are following. (i) In a finite volume $V$ and the finite 
momentum-cutoff at the scale $\pi/|{\bf a}|$, the total number of electron states in the 
energy-spectra (\ref{spectrum2}) and (\ref{spectrumh}) are finite and all 
these electron states of energy levels up to ${\mathcal E}_v$ are 
fully filled. (ii) The energy-spectrum (\ref{spectrum2}) is not degenerate, 
while the energy-spectrum (\ref{spectrumh}) is degenerate, and the total
numbers of electron states of both cases are the same. (iii) On the basis of 
quantum-mechanical fluctuations toward the lowest energy-state and the Pauli principle, 
when the external magnetic field $H$ is applied upon the semiconductor, the semi 
conductor reorganizes itself by fully filling all electron states according to the 
degenerate energy-spectrum (\ref{spectrumh}), instead of the non-degenerate 
one (\ref{spectrum2}). As a consequence, the semiconductor makes its total energy lower. 
As an analogy, the semiconductor with the energy-spectrum (\ref{spectrum2}) can be 
described as if a $N$-floors building, two rooms each floor and all rooms 
occupied by guests; while the semiconductor with the energy-spectrum (\ref{spectrumh}) 
a $M$-floors $(M<N)$ building, $2N/M$ rooms each floor and all rooms occupied by 
guests. The total numbers of rooms of two buildings are the same. Due to an 
external force, the $N$-floors building collapses to the $M$-floors building and 
the ``potential energy'' is reduced. 

In principle, when the external magnetic field is turned on, the semiconductor state gains energy, 
becomes energetically unstable and must decay to release the energy difference (\ref{delta2}) 
by quantum-mechanical fluctuations. As consequences, the possible phenomena and effects that 
could occur are following. (i) The semiconductor acts 
as a paramagnetic medium that effectively screens the strength of the external 
magnetic field $H$ to a smaller value $H'<H$ for the total energy-density being,
\begin{equation}
{1\over2}H'^2={1\over2}H^2-{\alpha \over3\pi}H^2\left |\frac{m_z}{m^*}\right|;
\quad
 H'=H\sqrt{1-
{2\alpha\over3\pi}\left |\frac{m_z}{m^*}\right|},
\label{hed}
\end{equation}
which should be possibly checked by appropriate experiments of precisely 
measuring the magnetic field strength.
(ii) Photons are spontaneously emitted analogously to the spontaneous photon 
emission for electrons at high-energy levels ``falling'' to low-energy levels in 
the atomic physics. In the following section, we compute rate and spectrum 
of spontaneous photon emission.

\section{Spontaneous photon emission}

We assume that the magnetic field ${\bf H}=H(t)\hat {\bf z}$ is spatially homogeneous 
in the dimension of semiconductor sample; and $H(t)$ is adiabatically turned on
\begin{equation}
H(t)=\Big\{\matrix{0,& t= t^-\rightarrow-\infty\cr H, & t= t^+\rightarrow+\infty}
\label{bt}
\end{equation}
in the time interval $\Delta\tau=t^+-t^-$. This adiabatic assumption means that the 
time variation (\ref{bt}) of magnetic field is much more slow than time evolution of Bloch electron 
states in semiconductors, which is characterized by the period 
$T^*=2\pi/\omega^*$ (\ref{spectrumh}) for motion in the 
$\hat {\bf x}-\hat {\bf y}$ plane, and by the time scale $\hbar/(m_zc^2)$ 
for motion in the $\hat {\bf z}$ direction. 


We adopt Eqs. (\ref{periodstate1},\ref{periodstateh}) and the assumption (\ref{periodstateapp})
for the ``initial'' Bloch electron state $\psi_i$ at $t\rightarrow -\infty, H=0$ and 
the ``final'' Bloch electron state $\psi_f$ at $t\rightarrow +\infty, H\not=0$.
With the normalization condition $\int d{\bf x}\psi^*\psi=1$, e.g. one particle per unit volume,
the initial Bloch electron state $\psi_i$ is given by, 
\begin{equation}
\psi_i ={1\over V^{1/2}}e^{\frac{i}{\hbar}[k_xx+\bar k_zz-{\mathcal E}_{\bf k}t]}.
\label{initial}
\end{equation}
Whereas the final Bloch electron state $\psi_f$ is,
\begin{equation}
\psi_f = {1\over (L_xL_z)^{1/2}}e^{\frac{i}{\hbar}[k'_xx+\bar k'_zz-
{\mathcal E}_n(k'_z)t]}\chi(y),
\label{final}
\end{equation}
where the expression of normalized function $\chi(y)$ is given in Ref. \cite{landau},   
\begin{equation}
\chi(y)= N_\chi^{1/2}e^{-\xi^2/2}H_n(\xi);\quad
N_\chi=\frac{(eH/c\hbar)^{1/2}}{2^n n!\pi^{1/2}};\quad  \xi=\left(\frac{eH}{\hbar c}\right)^{1/2}[y-\frac{ck'_x}{eH}]
\label{chi}
\end{equation}
and $H_n(\xi)$ is the Hermite polynomial. 
These initial and final Bloch electron states (\ref{initial}) and (\ref{final})
are associated with energy spectra (\ref{spectrum2}) and (\ref{spectrumh}) respectively.
Because Eqs. (\ref{spectrum2}) and (\ref{spectrumh}) possess 
the axial symmetry w.r.t. $\hat k_z$-direction in the ${\bf k}$-space, 
we chose $k_y=0$ in the initial state (\ref{initial}) to simplify calculations. The
initial momentum $k_z$ and final momentum $k'_z$ are the same, i.e. $k'_z=k_z$. This is can be understood 
by the semi classical equation of motion,
\begin{equation}
\frac{d{\bf k}}{d t} = \frac{e}{mc}{\bf k}\times {\bf H}.
\label{semi}
\end{equation}
for a magnetic field ${\bf H}$ along the $\hat {\bf z}$-direction.
  
The probability of spontaneous photon emission is related to 
$|J_{if}(q)|^2$, where the transition amplitude $J_{if}(q)$ 
from the ``initial'' $\psi_i$ (\ref{initial}) to the ``final'' state 
$\psi_f$ (\ref{final}) is given by (see for example \cite{LandauQuantumElectrodynamics})  
\begin{equation}
J_{if}(q)=-ie\int dtdxdydz \hskip0.1cm \psi^*_f\psi_i \left(\frac{c}{2\omega_qV}\right)^{1/2}
e^{\frac{i}{\hbar}(\omega_qt-{\bf q}\cdot{\bf r})},
\label{j1}
\end{equation}
where $\omega_q=c|{\bf q}|$ and ${\bf q}$ are photon's energy and momentum, and photon field is normalized by 
one photon energy $\omega_q$ crossing unit area per unit time. Based on the sizes of semiconductor sample are much larger than
the wavelengths of electrons states (\ref{initial},\ref{final}), i.e. $L_{x,y,z}\gg \hbar/k_{x,y,z}$ and $\Delta\tau \gg 2\pi/\omega^*$,
we approximate integrating ranges in Eq. (\ref{j1}) to be infinite.  
Using $\psi_i$ (\ref{initial}) 
and $\psi_f$ (\ref{final}), we integrate variables $t,x$ and $z$ in 
Eq.(\ref{j1}), giving rise to $\delta$-functions for energy-momentum conservations. 
Armed with Eq. (7.376) in \cite{gr}, we integrate variable $y$ in Eq. (\ref{j1}),
\begin{equation}
\int dye^{-\frac{i}{\hbar}q_yy}e^{-\xi^2/2}H_n(\xi)=(-i)^n({2\pi\hbar c\over eH})^{1/2}
e^{-\frac{i}{\hbar}q_y\left({ck'_x\over eH}\right)}H_n(\beta q_y)e^{-\beta^2q_y^2/2},
\label{j}
\end{equation}
where $\beta^2\equiv c/(eH\hbar)$. As results, Eq. (\ref{j1}) becomes,
\begin{eqnarray}
J_{if}(q)&=& (-i)^{n+1}e\left(\frac{cN_\chi}{2\omega_qVL_yL_x^2}\right)^{1/2}
({2\pi\hbar c\over eH})^{1/2}
e^{-\frac{i}{\hbar}q_y\left({ck'_x\over eH}\right)}H_n(\beta q_y)e^{-\beta^2q_y^2/2}\nonumber\\
&\cdot &(2\pi\hbar)\delta[{\mathcal E}_n(k'_z)-{\mathcal E}_{\bf k}+\omega_q ]
(2\pi\hbar)\delta(k_x-k'_x-q_x),
\label{jr}
\end{eqnarray}
where $q_z=0$ and $\omega_q=c(q_x^2+q_y^2)^{1/2}$ for $k_z=k_z'$ as discussed in Eq. (\ref{semi}). 
This indicates that emitted photons are in the direction perpendicular to the $\hat{\bf z}$-direction.    
The squared transition amplitude yields 
\begin{eqnarray}
|J_{if}(q)|^2 &=& e^2\left(\frac{cN_\chi}{2\omega_q VL_yL_x}\right)({2\pi\hbar c\over eH})
e^{-\beta^2q_y^2}H^2_n(\beta q_y)\Delta\tau  \nonumber\\
&\cdot&(2\pi\hbar)^2\delta[{\mathcal E}_n(k'_z)-{\mathcal E}_{\bf k}+\omega_q ]
\delta(k_x-k'_x-q_x).
\label{re0}
\end{eqnarray}
Using Eq. (\ref{intkn}), we integrate over all final states by 
summing $n$ with degeneracy ${\mathcal D}$ (\ref{degeneracy}), 
and obtain,
\begin{eqnarray}
\sum_f |J_{if}(q)|^2 &=& e^2\left(\frac{e^{-\beta^2q_y^2}}{\omega_q V}\right)
\left(\frac{|e|H}{\hbar c}\right)^{1/2}\Delta\tau c\cdot\nonumber\\
&&\sum_{n=0}^\infty{1\over2^n\pi^{1/2}n!}H^2_n(\beta q_y)(2\pi\hbar)
\delta[{\mathcal E}_n(k_z)-{\mathcal E}_{\bf k}+\omega_q ].
\label{re}
\end{eqnarray}
The $\delta$-function for the total energy-conservation can be written as  
\begin{equation}
(2\pi\hbar)\delta[{\mathcal E}_n(k_z)-{\mathcal E}_{\bf k}+\omega_q ] 
= (2\pi\hbar)\delta[(n+\lambda_0)\hbar\omega^* - \frac{k_x^2}{2m^*} - \omega_q ]
=\frac{1}{\omega^*}\delta_{n,n_\circ}
\label{delta}
\end{equation}
where the integer $n_\circ$ is determined by,
\begin{equation}
n_\circ+\lambda_0= \frac{1}{\hbar\omega^*} \left( \frac{k_x^2}{2m^*} + \omega_q  \right).
\label{nzero}
\end{equation} 
From Eqs. (\ref{re},\ref{nzero}), we obtain the probability of spontaneous photon emission 
from the whole volume of semiconductor sample per unit time
\begin{equation}
\frac{dN_\gamma}{dt} = {e^2\over 2^{n_\circ}\pi^{1/2} n_\circ !}
\left(\frac{eH}{\hbar c}\right)^{1/2}
\left(\frac{c}{\omega_q }\right)
e^{-\beta^2 q_y^2}H^2_{n_\circ}(\beta q_y).
\label{re1}
\end{equation}
Because of $q_z=0$ and the axial symmetry w.r.t $\hat {\bf z}$-direction in the phase space of photon's momenta ${\bf q}$, 
we can define the perpendicular momentum $|{\bf q}_\perp|\equiv q_y$ and photon energy $\omega_q=c|{\bf q}_\perp|$.
Eq. (\ref{re1}) the rate of spontaneous photon emission becomes, 
\begin{equation}
\frac{dN_\gamma}{dt}=  {e^2\over2^{n_\circ}\pi^{1/2} n_\circ !}
\left(\frac{eH}{\hbar c}\right)^{1/2}\left(\frac{c}{\omega_q}\right) 
\exp\left(-\frac{\omega_q^2}{eH\hbar c}\right)
H^2_{n_\circ}\left[\frac{\omega_q}{(eH\hbar c)^{1/2}}\right].
\label{re2}
\end{equation}
Actually, Eq. (\ref{re2}) is related to the number spectrum of spontaneous photon emission, and 
the corresponding energy spectrum is then given by
\begin{equation}
\frac{dE_\gamma}{dt}=\omega_q\frac{dN_\gamma}{dt}.
\label{snp}
\end{equation}

Since $\lambda_0\ge 0$, we take $n_\circ\ge 1$ in Eq. (\ref{re1}) to insure that Eq. (\ref{nzero})
is consistent. Eq. (\ref{re1}) shows that the photon energy $\omega_q$ has to be 
much less than the energy scale $(eH\hbar c)^{1/2}$, i.e. $\omega_q\ll (eH\hbar c)^{1/2}$, otherwise the 
rate (\ref{re2}) is exponentially suppressed. On the other hand, for larger $n^\circ$-values 
the rate (\ref{re2}) is proportional to 
\begin{equation}
\frac{1}{n^\circ !}\left(\frac{\omega_q}{(eH\hbar c)^{1/2}}\right )^{2n^\circ}\ll 1.
\label{small}
\end{equation}
Therefore, the leading contribution to the rates of spontaneous photon emission (\ref{re2}) and (\ref{snp}) are 
given by $n_\circ =1$, where the Hermite polynomial 
$H_1(x)=2x$, 
\begin{eqnarray}
\frac{d N_\gamma}{dt} &\simeq & {2e^2\over\pi^{1/2}}
\left(\frac{eH}{\hbar c}\right)^{1/2}\left(\frac{c}{\omega_q}\right)
\exp\left(-\frac{\omega_q^2}{eH\hbar c}\right)\frac{\omega^2_q}{eH\hbar c};
\label{re3}\\
\frac{d E_\gamma}{dt} &\simeq & {2e^2\over\pi^{1/2}}
\left(\frac{eH}{\hbar c}\right)^{1/2}\left(\frac{c}{\omega_q}\right)
\exp\left(-\frac{\omega_q^2}{eH\hbar c}\right)\frac{\omega^3_q}{eH\hbar c}.
\label{re3e}
\end{eqnarray}
This number (energy) spectrum shows $\omega_q$ ($\omega_q^2$) dependence 
in the low-energy region and $\exp[-\omega_q^2/(eH\hbar c)]$ dependence 
in the high energy region. The number spectrum (\ref{re3}) is maximum at $\omega_q=\omega_q^{\rm max}=(eH\hbar c/2)^{1/2}$. 
The energy scale $\sqrt{eH\hbar c}\simeq 0.244$eV for $H=10^5$ Gauss achieved in 
the laboratory today \cite{magnet}. 
This indicates that most photons emitted are in the infrared region for $H< 10^5$ Gauss.  

In Eqs.~(\ref{re3},\ref{re3e}), integrating over photon's energy $\int d\omega_q/(2\pi\hbar c)$, 
we obtain the total number and energy of spontaneous emitted photons 
per unit time from the whole volume of semiconductor,
\begin{eqnarray}
\frac{dN_{\gamma}}{dt} &\simeq & 
\frac{2\alpha}{\pi^{1/2}}\frac{(eH\hbar c)^{1/2}}{\hbar};
\label{np1}\\
\frac{dE_\gamma}{dt} &\simeq &\alpha
\frac{(eH\hbar c)^{1/2}}{\hbar} (eH\hbar c)^{1/2}.
\label{snp1}
\end{eqnarray}
Note that we have to take into account contributions from all different energy bands 
${\mathcal N}$ of semiconductors. 

There are two degenerate valence band maxima, both located at ${\bf k}=0$, in both Silicon and Germanium.  
These degenerate valence band maxima are spherically symmetric ($m^*=m_z$) in the valid quadratic approximation
(\ref{spectrum1}). The effective masses $m^*$ are $0.49m$ and $0.1m$ (Silicon); 
$0.28m$ and $0.044m$ (Germanium) \cite{solidbook}. In order for an estimation, 
we take $m^*=0.1mc^2$ and $\sqrt{eH\hbar c}\simeq 0.244$eV for $H=10^5$Gauss. 
The characteristic time $T^*=2\pi/\omega^*\simeq 3.3\cdot 10^{-7}$sec and $\hbar/(m^*c^2)=1.3\cdot 10^{-20}$ sec. The rates
Eqs. (\ref{np1},\ref{snp1}) give  
\begin{equation}
\frac{dN_{\gamma}}{dt} \simeq  3.0\cdot 10^{12} {\rm sec}^{-1};\quad 
\frac{dE_\gamma}{dt} \simeq 6.6\cdot 10^{11}{\rm eV}{\rm sec}^{-1}.
\label{estimation}
\end{equation}
One should be able to detect such photon emission in the direction perpendicular to the direction of 
magnetic field $H$, and measure the number and energy spectra (\ref{np1},\ref{snp1}), as well as  
rates (\ref{estimation}).

\section{Some remarks}

Strictly speaking, the rate and spectrum of spontaneous photon emission 
should depend on the way of turning on the magnetic field $H(t)$ in the time interval $\Delta\tau$. 
Our calculations of the rate of spontaneous photon emission base on the adiabatic assumption 
$\Delta \tau \gg T^*=2\pi/\omega^*=3.3\cdot 10^{-7}$sec., which can be easily satisfied in experiments.
In addition, a slowly time-varying magnetic field $H(t)$ leads to a slowing time-varying electromagnetic fields, 
inducing a background of soft photons. These soft photons however have a typical energy $\hbar/\Delta\tau$,
which is much smaller than the typical energy $\sqrt{eH\hbar c}$ of spontaneous emitted photons
discussed in above. Therefore this photon background should not be a problem for 
experiments detecting spontaneous emitted photons. Beside, experiments should be performed at low temperature to
reduce the background of thermal photons, which is about $10^8/{\rm cm}^3$ at the room temperature.     
      
All these discussions and calculations in the present article can be also applied to the energy spectrum of 
metals at low temperature. At low temperature, the Fermi surfaces of energy spectrum of metals very sharp, 
i.e. very few electrons are exited on the Fermi surface. The skin and screening effect can be avoided and 
the magnetic field can penetrate whole volume of metals. The effect of spontaneous photon emission 
should be produced by the same mechanism as in the semiconductor case.  

Considering an periodic oscillating magnetic field $H(t)=H_0\cos(\omega_H t)$ with a period $T_H=2\pi/\omega_H\gg T^*$, 
we can still apply the adiabatic assumption. In the phase that the magnetic field increases from $0$ to maximum $H_0$,
the spontaneous photon emission occur as discussed. If these photons are not kept by either a large opacity or a cavity, they
must stream away with energy. In the phase that the magnetic field decreases from the maximum $H_0$ to $0$ and the energy
spectrum of semiconductors (or metals at low temperature) has to be changed from a degenerate spectrum (\ref{spectrumh}) to a 
non-degenerate spectrum (\ref{spectrum1}), we speculate that a semiconductor 
should absorb thermal photons from its environment to make such a transition. As a consequence, an periodic oscillating magnetic 
field $H(t)$ acting on semiconductors results in a cooling process to decrease temperature. 

Such effect of spontaneous photon emission should also be observed in astrophysics 
processes of a large variation of strength of magnetic fields, e.g., supernova explosions and 
neutron stars. The energy scale $\sqrt{eH\hbar c}\simeq 24$KeV for $H\simeq 10^{15}$ Gauss around 
neutron stars, which can be probably considered as a conductor or semiconductor.
It is worthwhile to mention that such effect of spontaneous photon emission  
could account for the anomalous X-ray pulsar \cite{xray}.

We expect sophisticate experiments in near future to verify spontaneous photon emission 
(\ref{re3},\ref{re3e}) and screening effect (\ref{hed}), when semiconductors are in a varying magnetic field. 
This is important for understanding energy spectra of semiconductors 
and any possibly prospective applications.

I thank Prof. H. Kleinert who suggest me to study the case of semiconductors.






\begin{thebibliography}{99}

\bibitem{xue}
S.-S. Xue , Phys. Lett. B508 (2001) 211;  Phys. Rev. D68. (2003) 013004.

\bibitem{solidbook}
Neil W. Ashcroft and N. David Mermin, ''Solid State Physics'' HRW International Editions,
CBS Publishing Asia LTD. (1975)

\bibitem{landau}
L.D. Landau, and E.M. Lifshitz, {\it Quantum Mechanics} (2nd ed.) Addition-Wesley,
Reading, Mass., 1965 (1965) 424-426.

\bibitem{magnet}
see ``http://www.nhmfl.gov/magtech/lhfs/index.html''. 

\bibitem{LandauQuantumElectrodynamics}
V.B. Beresretskii, E.M. Lifshitz and L. L.P. Pitaevskii, {\it Quantum Electrodynamics}, 2nd ed. (Pergamon Press,
New York, 1982),

\bibitem{gr}
I.S.~Gradshteyn and I.M.~Ryzhik, ``Table of integrals, series, and products'' 
(1965) New York Academic Press.

\bibitem{xray}
G.L.~Israel, T.~Oosterbroek, L.~Stella, S.~Campana, S.~Mereghetti and A.N.~Parmar
astro-ph/0108506, ApJ {\bf 560}, L65 (2001).

\end{thebibliography}
\end{document}